\documentstyle[12pt]{article}
\begin{document}

\newcommand{\appsection}{\addtocounter{section}{1}
	   \setcounter{equation}{0}\section*{Appendix\Alph{section}}}
\renewcommand{\theequation}{\arabic{equation}}

\newcommand{\be}{\begin{equation}}
\newcommand{\ee}{\end{equation}}
\newcommand{\bdm}{\begin{displaymath}}
\newcommand{\edm}{\end{displaymath}}
\newcommand{\beann}{\begin{eqnarray*}}
\newcommand{\eeann}{\end{eqnarray*}}
\newcommand{\bea}{\begin{eqnarray}}
\newcommand{\eea}{\end{eqnarray}}

\newcommand{\nn}{\nonumber \\}

\begin{titlepage}

\def\mytoday#1{{ } \ifcase\month \or
 January\or February\or March\or April\or May\or June\or
 July\or August\or September\or October\or November\or December\fi
 \space \number\year}
\noindent
\hspace*{11cm} BUTP--94/25\\
\vspace*{1cm}
\begin{center}
{\LARGE On the Three-Point Vertex of Hadron Interaction with External Fields
in QCD Sum Rules}

\vspace{2cm}

B.L. Ioffe
\\
Institute of Theoretical and Experimental Physics \\
B. Cheremushkinskaya 25, RU-117259 Moscow, Russia \\
and \\
Institute for Theoretical Physics \\
University of Bern \\
Sidlerstrasse 5, CH-3012 Bern, Switzerland\\
e-mail:ioffe@vxdesy.desy.de, ioffe@vxitep.itep.ru
\vspace{1.5cm}

\mytoday \\ \vspace*{1cm}

\nopagebreak[4]

\begin{abstract}
The calculation method of the hadron interaction vertex with constant external
fields in QCD sum rules is discussed. The representation of the polarization
operator in terms of physical states contributions is considered and the most
suitable form of this representation is suggested. The estimates of
uncertainties in the previous calculations of hadron interaction vertices with
external fields are given.

\end{abstract}

\end{center}

\end{titlepage}

\section{Introduction}

The QCD sum rule calculations of hadron interaction vertices with constant
external fields are widely used for determination of the hadron static
properties. In this way the proton, neutron and hyperon magnetic moments [1-5]
, the axial nucleon and hyperon $\beta$-decay coupling constatns [6-8], the
$\pi NN$, $\pi N \Delta$ coupling constants [9] and many other static hadronic
characteristics were found. More recently the same method was used to determine
the moments of the structure functions [10] and higher twist corrections to
Gross-Llewellyn and Bjorken sum rules for unpolarized [11] and polarized [12]
deep
inelastic lepton-nucleon scattering.

 I will remind the main features ot the
method. The term of quark interaction with external field is added to QCD
Lagrangian. This field may be a constant electromagnetic field in the case of
magnetic moments calculations, a fictititious constant axial potential in the
case of $g_A$ and higher twist correction to polarized Bjorken sum rule etc.
The polarization operator $\Pi(p^2)$ of currents with the quantum numbers of
hadrons in view is considered. It is supposed that $p^2$ is negative and $- p^2
\gg R_c^{-2}$, where $R_c$ is the confinement radius, and few terms in the
operator product expansion (OPE) of $\Pi(p^2)$ are calculated. For quarks the
Dirac equation in the external field is written and only the terms linear in
this field are retained. The term in the polarization operator linear in the
external source is given by
\be
\Pi(p^2) = i^2 \int d^4x e^{ipx} <0 \vert T \{ \eta(x), \int j(z) d^4z,
\bar{\eta} (0) \} \vert 0> S,
\ee
where $S$ is the external field, $\eta$, $\bar{\eta}$ are the currents with the
quantum numbers of the hadron, whose interaction vertex with external current
$j$, we would like to determine. (E.g. for the case of baryons, $\eta(x)$ is
proportional to the product of three quark fields, $\eta \sim \varepsilon^{abc}
q^aq^bq^c$, where $a,b,c$ are colour indices.) The important ingredient of the
calculation is the account of induced  by the external vield vacuum
expectation
values in the OPE. In this way the linear in the external field term in the
polarization operator, proportional to the vertex function with zero momentum
transfer $\Gamma (p^2) \equiv \Gamma(p^2,p^2, 0)$,  $\Pi = \Gamma S$ is
calculated by OPE in QCD. On the other side using dispersion relation,
$\Pi(p^2)$ is represented through contributions of hadronic states. Among these
the contribution of the lowest state in the given channel is of interest and
must be separated. If the contributions of exited states can be reliably
estimated and are small, then by equating of two representations of $\Pi(p^2)$
the desirable hadron interaction vertex with constant external field can be
found.

In carrying out this program it is essential to represent correctly the
hadronic contributions to $\Pi(p^2)$, or what is equivalent, to $\Gamma(p^2)$,
accounting for all possible terms. The omission of some terms may result in
underestimation of possible errors in the values of hadronic coupling constants
with external fields and even in completely wrong results. Unfortunately, not
in all such calculations the general form of hadronic spectra was used and the
errors due to this, were estimated. The goal of this paper is to discuss the
general form of hadronic contributions to $\Pi(p^2)$ in the external field (or
$\Gamma(p^2)$) and to suggest the most suitable form for representation of
excited states. For few examples the uncertainties in the determination of
hadronic coupling constants with external fields arising from contributions of
excited states will be estimated.

\section{Dispersion representation of the vertex function}

In order to get the dispersion representation of the polarization operator in
the external field $\Pi (p^2)$ or the vertex function at zero momentum
transfer $\Gamma(p^2,p^2, 0)$ it is convenient to start
from the case, when the momentum transfer $q=p_2-p_1$ is not zero, but $q^2$ is
small and negative. (All what follows refers to the coefficient function at any
Lorentz tensor structure.) The general double dispersion representation in
variables $p^2_1,p^2_2$ of $\Gamma(p^2_1, p^2_2, q^2)$ at fixed $q^2 <0$ has
the form (see, e.g. [2])
\be
\Gamma(p^2_1, p^2_2, q^2) = \int^{\infty}_{0} \int^{\infty}_{0} \frac{\rho(s_1,
s_2, q^2)}{(s_1 -p^2_1)(s_2-p^2_2)} ds_1 \; ds_2 +P(p^2_1) f(p^2_2, q^2)
+ P(p^2_2) f(p^2_1, q^2),
\ee
where $P(p^2)$ is a polynomial. (It can be shown that at $q^2 <0$ there are no
anomalous thresholds.) For simplicity it is assumed that the currents $\eta$
and $\bar{\eta}$ are Hermite conjgate what correspond to the diagonal matrix
element over hadronic states. In this case $\rho(s_1, s_{2},q^2)$ is
symmetric in $s_1, s_2$ and the functions $f$ are the same in the
2$^{\mbox{nd}}$ and $3^{\mbox{rd}}$ terms in the r.h.s. of (2). The second and
the third terms in the r.h.s of (2) plays the role of subtraction functions in
the double dispersion relation (2). The function $f(p^2,q^2)$ may be
represented by one-variable dispersion relation in $p^2$.

It is clear that the dispersion representation (2) holds also in the limit $q
\rightarrow 0$, $p^2_2 \rightarrow p^2_1 \equiv p^2$, where $\Gamma(p^2)
\equiv
\Gamma(p^2,p^2, 0)$ is a function of one variable $p^2$. At first sight
it seems that one variable dispersion relation can be written for
$\Gamma(p^2)$. Indeed, decomposing the denominator in (2), we can write
\be
\int^{\infty}_{0} ds_1 \int^{\infty}_{0} ds_2 \frac{\rho (s_1, s_2)}{(s_1 -
p^2)(s_2-p^2)} = \int \int \frac{\rho (s_1,s_2)}{s_1-s_2} ds_1 ds_2 \left(
\frac{1}{s_2 -p^2} -\frac{1}{s_1 - p^2} \right ).
\ee
In the first (second) term in the r.h.s. of (3) the integration over $s_1(s_2)$
can be performed and the result has the form of one-variable dispersion
relation. Such transformation is, however, misleading, because, in general, the
integrals
\be
\int ds_1 \frac{\rho(s_1, s_2)}{s_1 - s_2} = - \int ds_2 \frac{\rho(s_1,
s_2)}{s_1-s_2}
\ee
are ultravioletely divergent. This ultraviolet divergence cannot be cured by
subtractions in one-variable dispersion relation: only the subtractions in the
double dispersion representation (2) may be used. It is evident that the
procedures, killing the subtraction terms and leading to fast converging of
dispersion integrals in standard one-variable dispersion representations, like
Borel transformation in $p^2$ do not help here.

Let us consider two examples. The first corresponds to the determination of
nucleon magnetic moments. In this case the current $\eta$ in (1) is the quark
current with nucleon quantum numbers $\eta \sim \varepsilon^{abc} q^aq^b q^c$
and the current $j$ in (2) is the electromagnetic current. The simple loop
diagram -- the contribution of unit operator in OPE -- is shown in Fig. 1. It
is clear that the spectral
function $\rho(s_1, s_2)$ in (2), corresponding to the diagram of Fig. 1 is
proportional to $\delta (s_1 -s_2)$. The separation of the chirality conserving
structure results in the statement that the dimension of $\rho$ is equal to 2
(see [2,3]). So the general form of $\rho(s_1, s_2)$ in (2) is
\be
\rho (s_1, s_2) = a s_1 s_2 \delta (s_1 - s_2),
\ee
where $a$ is a constant. The substitution of (5) into (2) gives for the first
term in the r.h.s. of (2) at $p^2_1 = p^2 \equiv p^2$
\be
\Gamma (p^2) = a \int^{\infty}_{0} \frac{s^2_1 ds_1}{(s_1 - p^2)^2}
\ee
In this simple example the dispersion representation is reduced to
one-variable dispersion relation, but with the square of $(s_1 - p^2)$ in the
denominator. Of course, by integrating by parts (6) may be transformed to the
standard dispersion representation. However, the boundary term, arising at such
transformation must be accounted; it does not vanish even after application of
Borel transformation. This means that, even in this simplest case, the
representation (2) is not equivalent to one-variable dispersion relation.

The second example corresponds to determination of twist 4 correction to the
Bjorken sum rule for polarized deep inelastic electron-nucleon scattering [12].
Here the external current $j$ in (1) is given by
\be
U_\mu = \frac{1}{2} \bar{q} g \varepsilon_{\mu \nu \lambda \sigma} G_{\lambda
\sigma}^a \lambda^a q,
\ee
where $G_{\mu \nu}^a$ is the gluonic field strength tensor. An example of
the
bare loop diagram is shown in Fig. 2. In this case, unlike the previous one,
$\rho(s_1, s_2)$ is not proportional to $\delta (s_1 - s_2)$. This stems from
the fact that in the discontinuity over $p^2_1$ at $q \neq 0$ and $p^2_1 \neq
p^2_2$ only the left-hand part of the diagram Fig. 2 is touched and the loop
integration in the right-hand part still persists.  For the selected in ref.
[12] tensor structure $\rho(s_1, s_2)$ has dimension 4 and is proportional to
$s_1 s_2$. We see that in this example the general form of dispersion
representation (2) must be used in the limit
\be
q^2 \rightarrow 0, p^2_1 \rightarrow p^2_2 = p^2.
\ee

Let us represent $\Gamma(p^2, p^2;0)$ in terms of contributions of hadronic
states, using (2) and separating the contribution of the lowest hadronic state
in the channels with momentum $p$. Consider the first term in the r.h.s of (2)
at $q^2 = 0, p^2_1 = p^2_2 = p^2$. As is seen from Fig. 3 it is convenient to
devide the whole integration region in $s_1,s_2$ into three domains: I) $0 <
s_1 < W^2$, $0 < s_2 < W^2$; II) $0 < s_1 < W^2$, $W^2 < s_2 < \infty$; $W^2 <
s_1 < \infty$, $0 < s_2 < W^2$; III) $W^2 < s_1 < \infty$, $W^2 < s_2 <
\infty$. Adopt the standard in QCD sum rule model of hadronic spectrum: the
lowest hadronic state plus continuum, starting from some threshold $W^2$. Then
in the domain I only the lowest hadronic state $h$ contributes and
\be
\rho (s_1, s_2) = G \lambda^2 \delta (s_1 - m^2) \delta (s_2 - m^2)
\ee
where $m$ is the mass of this state, $\lambda$ is the transition constant of
the hadron in the current $\eta$. (For $h$-baryon $ <B \vert \bar{\eta} \vert
0>= \lambda_B \bar{v}_B$ where $v_B$ is the baryon spinor.) $G$ is the coupling
constant
of the hadron with external field, which we would like to determine from the
sum rule. In the domain III the higher order terms in OPE may be neglected and
the contribution of hadronic states is with a good accuracy equal to the
contribution of the bare quark loop (like Figs. 2 or 3) with perturbative
corrections. The further application of Borel transformation in $p^2$
essentially suppresses this contribution.

The consideration of the domain II contribution is the most troublesome and
requires an additional hypothesis. Assume, using the duality arguments, that in
this domain also, the contribution of hadronic states is approximately equal to
the contribution of the bare quark loop. The accuracy of this approximation may
be improved by subtraction from each strip of domain II of the lowest hadronic
state contributions, proportional to $\delta (s_1 - m^2)$ or $\delta (s_2 -
m^2)$. Terms of the latter type also persist in the functions $f(p^2_1),
f(p^2_2)$ in (2). They correspond to the process when the current $\bar{\eta}$
produces the hadron $h$ from the vacuum and under the action of the
external current $j$ the transition to excited state $ h \rightarrow h^\star$
occurs or vice versa (Fig. 4). At $p^2_1 = p^2_2 = p^2$ these contributions
have the form
\be
\int^{\infty}_{W^2} \frac{b (s) ds}{(p^2-m^2) (s-p^2)}
\ee
with some unknown function $b(s)$. The term (10) will be accounted separately
in the r.h.s. of (2). I stress that the term (10) must be added
to the r.h.s. of (2) independently
of the form of bare loop contribution $\rho(s_1, s_2)$. Even if $\rho(s_1,s_2)
= 0$, when the OPE for the vertex function $\Gamma(p^2, p^2, 0)$ with zero
momentum transfer starts from condensate terms -- the term (10) may persist
(the example of such a situation will be given in Sec. 3).
(10) may be written as
\be
\int^{\infty}_{W^2} ds b(s) \left ( \frac{1}{p^2 - m^2} +
\frac{1}{s-p^2}\right ) \frac{1}{s-m^2}..
\ee
The functions $f(p^2)$ in (2) can be represented by dispersion relation as
\be
f(p^2) = \int^{\infty}_{0} \frac{d(s)}{s-p^2} ds
\ee
The integration domain in (12) may be also devided into two parts: $0<s<W^2$
and $W^2 <s< \infty$. According
to our model the contribution of the first part is approximated by $h$-state
contribution, the second one by continuum. These two parts look like the
contributions of the first and the second terms in the bracket in (11). The
first term in (11), which after Borel transformation is not suppressed in
comparison with the main double pole term, arising from (7), as a rule is
accounted in the calculations (see, e.g. [1-8, 10-12]) as an unknown parameter,
determined from the same sum rule. The second term in (11), suppressed in
comparison with the first one after Borel transformation, is traditionally
neglected.

Now we can formulate the  recipee, how the sum rule can be written. At the
phenomenological side -- the r.h.s of the sum rule -- there is the contribution
of the lowest hadronic state $h$ and the unknown term (11), corresponding to
nondiagonal transition $h \rightarrow h^\star$ in the presence of external
field:
\be
\frac{\lambda^2 G}{(p^2-m^2)^2} + \int^{\infty}_{W^2} ds b(s) \frac{1}{s-m^2}
\left (\frac{1}{p^2-m^2} + \frac{\alpha(s)}{s-p^2} \right )
\ee
(The coefficient $\alpha$ reflects the possibility that in the function $f$
the
ratio of terms, proportional to $(p^2-m^2)^{-1}$ and $(s-p^2)^{-1}$ may differ
from 1 as it takes place in (11).) The contribution of continuum,
corresponding
to the bare loop (or also to the higher order terms in OPE, if their
discountinuity does not vanish at $s \rightarrow \infty$) is transferred to the
l.h.s. of the sum rule. Here it is cancelled by the bare loop contribution from
the same domain of integration. As a result in the double dispersion
representation of the bare loop the domain of integration over $s_1,s_2$ is
restricted to $0 < s_1, s_2 <W^2$. Finally, apply the Borel transformation in
$p^2$ to both sides of the sum rule. In the l.h.s. -- QCD side -- the
contribution of the bare loop has the form
\bdm
\int^{W^2}_{0} ds_1 \int^{W^2}_{0} ds_2 \rho (s_1, s_2) \frac{1}{s_1-s_2}
\left[e^{-s_2/M^2}- e^{-s_1/M^2}\right]
\edm
\be
= 2P \int^{W^2}_{0} ds_2 \int^{W^2}_{0}
ds_1 \frac{\rho (s_1,s_2)}{s_1-s_2} e^{-s_2/M^2},
\ee
where $P$ means the principal value and the symmetry of $\rho(s_1,s_2)$ was
used. The r.h.s. of the sum rule is equal;
\be
G \frac{\lambda^2}{M^2} e^{-m^2/M^2}- Ae^{-m^2/M^2}+e^{-m^2/M^2}
\int^{\infty}_{W^2} ds b(s) \frac{\alpha(s)}{s-m^2} ~\exp[-(s-m^2)/M^2]
\ee
where
\be
A = \int^{\infty}_{W^2} ds \frac{b(s)}{s-m^2}
\ee

Two remarks in connection with eqs. (14), (15) are necessary. If the
discontinuity $\rho(s_1, s_2)$ of the bare loop is proportional to
$\delta(s_1-s_2)$, $\rho (s_1, s_2) = \rho(s_1) \delta(s_1-s_2)$ like in the
diagram Fig. 1, then eq. (14) reduces to
\be
\frac{1}{M^2} \int^{W^2}_{0} ds_1 e^{-s_1/M^2} \rho(s_1).
\ee
In this case at $W^2 \gg M^2$ the continuum contribution is suppressed
exponentially and the dependence on the value of continuum threshold $W^2$ is
weak. It, however, $\rho(s_1,s_2)$ has no such form and is a polynomial in
$s_1,s_2$, like in the diagram Fig. 2, then, as can be be seen from (14), the
bare loop diagram contribution has a powerlike dependence on $W^2$.
In this aspect the QCD sum rule calculation lost a part of its advantages in
comparison with finite energy sum rules.

In the case, when the double discontinuity of the bare loop diagram
$\rho(s_1,s_2)$ is proportional to $\delta(s_1-s_2)$, this form will be absent
in the radiation correction terms. Here (17) is invalid and more general
expression (14) must be used. This will result to appearance of
$\ln(W^2/-p^2)$
in the final answer -- $W^2$ plays the role of ultraviolet cut off.

The necessity to account separately the second term in (15) in the r.h.s. of
(15) not suppressed after Borel transformation comparing with the first term,
was stressed in the early treatment of the problem in view [1-3]. In [1-3] it
was suggested to separate this unknown term from proportional to $G$ term of
interest by their different preexponent $M^2$ dependence. This program was
successfully realized in the most cases of hadron coupling constants with
external fields determinations. However, the last term in (15), was omitted in
all calculations, following the first ones [1-3]. This term is suppressed in
comparison with the accounted second term in (15) by a factor, less than
$e^{-(W^2-m^2)/M^2}$. In the most cases this factor is of order $e^{-1.5}
\approx
1/4$. Therefore, if $A \ll G\lambda^2/M^2$ this term can be safely neglected.
But at $A \sim G
\lambda^2/M^2$ the nonaccounted term may
deteriorate the accuracy of the results.

\section{Few examples}

\subsection{Quark mass term}

Consider the matrix element
\be
H = < p \vert \bar{u}u - \bar{d}d \vert p> /2m_p
\ee
over the proton state $\vert p >$. Here $u$ and $d$ are the fields of $u$ and
$d$-quarks. This matrix element was studied recently [13] by the QCD sum rule
technique. On the other side due to the Hellman-Feynman theorem [14,15] $H$ is
related [13] to the part of the neutron-proton mass difference, arising from
the quark mass difference $\mu = m_d-m_u$
\be
(m_n - m_p)_\mu = \mu H
\ee
in the linear over $\mu$ approximation.
Therefore $H$ may be calculated in two ways in the QCD sum rule approach: 1)
considering three point hadron vertex for interaction with external constant
$\bar{u}u-\bar{d}d$ quark field, as was done in [13]; 2) considering the quark
mass dependence of proton and neutron mass in the framework of nucleon mass
calculation in QCD sum rule, as was done in [16,17]. In the latter case the
quantity $(m_n-m_p)_\mu$ -- the l.h.s. of (19) is calculated. It is evident,
that since both calculations are based on the same physical ideas and using the
same technique, they must be in on-to-one correspondence. The comparison of the
QCD sides of the sum rules, found in refs. [13] and [17] shows that it is
indeed the case: the QCD sides of the sum rules for $(m_n-m_p)_\mu$, obtained
in [17], identically conincide with QCD sides of the sum rules for $\mu H$,
found in [13] \footnote{There was an error in ref. [17] in the coefficient in
front of the square of quark condensate term in chirality violating structure,
instead of 4/3 it must be 2/3. This error was noticed in ref. [13]. The
results
of ref. [17] are unaltered.}. However, the phenomenological sides of the sum
rules in refs. [13] and [17] are different.

The origin of various terms in the phenomenological part of QCD sum rules in
the calculation, when the quark mass dependence of the nucleon mass was
studied, can be easily understood. Consider, for example, the chirality
conserving sum rule for nucleon mass. After Borel transformation the
phenomenological part of the sum rule take the form
\be
\lambda^2_N e^{-m^2/M^2} + c \int^{\infty}_{W^2} e^{-s/M^2}s^2 ds,
\ee
where $c$ is some QCD calculated numerical coefficient. We are interested in
the dependence of (20) on the quark mass difference $\mu$ in the linear in
$\mu$ approximation. So, differentiate (20) over $\mu$. We have
\be
- \lambda^2_N \frac{\delta m^2}{M^2} e^{-m^2/M^2} + \delta \lambda^2_N
e^{-m^2/M^2} - c\delta W^2 e^{-W^2/M^2} W^4,
\ee
where $\delta W^2$ is the
variation of continuum threshold. Compare (21) with the phenomenoloigcal
side of the sum rule for the vertex $H$, which has the form (15). If the
last term in (15) is neglected, as was done in [13], then we have
\be
\frac{H}{M^2} e^{-m^2/M^2} - A e^{-m^2/M^2}.
\ee
The first term in (22)
corresponds to the first proportional to $\delta m^2$, term in (21) and
completely agrees with the Feynman-Hellman-theorem. The second term in (22),
proportional to unknown constant $A$, corresponds to the second term in
(21), where $\delta \lambda^2_N$ is also unknown and as well as $A$ is
determined from the same sum rule. But nothing corresponds to the last term
in (21) in eq. (22). This term is proportional to the variation of continuum
threshold with the quark mass. It is evident that the continuum threshold as
all hadronic masses is varying with the quark masses. Therefore, the last
term in (21) generally is non-zero. The contradiction between eqs. (21),
(22) disappears if the last term in (15) is accounted. The functional $M^2$
dependence of this term is the same as one of the last term in (21), since
at $W^2 >> M^2$ the integral in (15) is concentrated at low limit

The requirement of one-to-one correspondence of two sum rules -- the sum rule
for vertex function and the sum rule determining the mass difference, following
from the Hellman-Feynman-theorem, results in a specific form of the last term
in (15), if the standard form of hadronic spectrum is used in the sum rules
for mass difference. But this is not a general case: generally the last term in
(15) is arbitrary.

We see on this example that the last term in (15) must be accounted. This
introduces an uncertainty in the calculation of three point vertex functions in
the QCD sum rule approach.

Consider now another example of the quark mass term contribution -- the matrix
element
\be
H_V = < V \vert \bar{q}q \vert V > / 2m_V
\ee
where $V$ is the vector meson, built from $\bar{q}q$ pair, say
$\varphi$-meson.
The Hellman-Feynman-theorem relates $H_\varphi$ to $\varphi - \omega$ (or
$\varphi - \rho$) mass difference
\be
(m_\varphi - m_\omega) = m_s H_\varphi
\ee
Due to chiral invariance the OPE for polarization operator in the chirality
violating external field $\bar{q}q$ starts from the operator $m_\varphi
\bar{q}q$ and the first term in the OPE corresponds to the diagram Fig. 5. The
dimension of this operator is 4, its contribution vanishes at large $s$ and the
structure function $\rho(s_1,s_2)$ in (2) is zero. However, the second term
in eq. (15) is nonzero. This is evident, if we repeat the same consideration
for meson mass determination, which were done for nucleon mass and led to
(20), (21). Again, the requirement that the Hellman-Feynman-theorem holds,
leads to the statement that the last term in (15) must be retained. There
are no reasons to neglect it, since we are sure that the masses of excited
states in $\varphi$-channel are heavier than in $\omega$-channel, and, as a
consequence, the term proportional to $\delta W^2$ in (21) exists. This
example demonstrates the necessity to account the term with nondiagonal
transitions $h \rightarrow h^\star$ in its general form, given by the last
two terms in eq. (15) even if the bare loop diagram is absent.

\subsection{Nucleon magnetic moments and the axial coupling constants $g_A$}

In these cases, [1-8] the bare loop diagram is proportional to $\delta (s_1
-s_2)$
and the correct form of continuum contribution corresponds to $(s_1-p^2)^2$ in
the denominator, as in eq. (6). This form was used in ref. [2] (see Note Added
in Proof) and in refs [6-8]. The nondiagonal constant term (17) corresponding
to $N \rightarrow N^\star$ transitions in the external field -- the second term
in (15) -- was accounted in the calculations of nucleon and hyperon magnetic
moments and $g_A$ [1-8], but the last term in (15) was neglected. Let us
estimate, how its account can influence the results.

In the case of nucleon magnetic moments the constants $A_{p,n}$ and $B_{p,n}$,
corresponding to the constant $A$ in (15) were found from the sum rules in [2]
(see Note Added in Proof). Taking these values and assuming that the third term
in (15) is suppressed in comparison with the second one by the factor $\exp [-
(W^2-m^2)/M^2]$, $W^2 = 2.3 \mbox{GeV}^2, M^2 \approx 1 \mbox{GeV}^2$, we find
that the uncertainty in the proton and neutron magnetic moments, arising from
this source is $\vert \Delta \mu_p \vert \approx \vert \Delta \mu_n \vert
\approx 0.015$, much below the overall uncertainty, estimated in [2] -- $\vert
\Delta \mu_{p,n} \vert / \vert \mu_{p,n}\vert \approx 10\%$.

The similar procedure can be performed in the case of isovector nucleon axial
coupling constant $g_A$ determined in [6] and octet axial coupling constant
$g_A^s$ found in [8]. The result is the same: the uncertainty, arising due to
nonaccounted in [6,8] terms, corresponding to the last term in (15) are much
less (about 3-4 times) than the extimated overall error.

\subsection{Twist four correction to Bjorken sum rule for polarized deep
inelastic lepton-nucleon scattering}

As shown in [18] the twist four correction to Bjorken sum rule for deep
inelastic polarized lepton-nucleon scattering is expressed through the matrix
element over nucleon state of the operator
\be
U_\mu = \bar{u} g \tilde{G}^a_{\mu \nu} \gamma_\nu \frac{1}{2} \lambda^a u -
\bar{d}g \tilde{G}^a_{\mu \nu} \gamma_\nu \frac{1}{2} \lambda^a d,
\ee
where $\tilde{G}^a_{\mu \nu} = (1/2) \varepsilon_{\mu \nu \lambda \sigma}
G^a_{\lambda \sigma}$ is the dual gluonic field strength tensor. The
calculation of the matrix element $< p \vert U_\mu \vert p > \equiv s_\mu <<U
>>$ in the framework of QCD sum rules was performed by Balitsky, Brown and
Kolesnichenko [12]. The bare loop diagram for this case is shown in Fig. 2. In
[12] this diagram was calculated, by introducing an ultraviolet cut off
$\mu^2$. It was found that for the chosen Lorentz structure the singular in
$p^2$ term is proportional to $p^4 \ln^2 (\mu^2/ - p^2)$-. Such cut off
dependence reflects the fact that the spectral function $\rho(s_1, s_2)$ in eq.
(2) is not proportional to $\delta$-function. The logarithmn square dependence
of $\Gamma(p^2)$ on the cut-off cannot be removed by Borel transformation.
For this reasons, in order to obtain physical results the authors of ref. [12]
considered various values of $\mu^2$ in the interval $0 \vert < \mu^2 < 1
\mbox{GeV}^2$ and included uncertainties arising from this procedure into the
error. From the presented above point of view such an approach is not
legitimate. In this case $\rho(s_1, s_2)$ is proportional to $s_1 s_2$:
\be
\rho (s_1,s_2) = b s_1 s_2
\ee
where $b$ is a constant. In the model of hadronic spectrum accepted in Sec.2,
we have after Borel transofrmation and using eq. (14):
\bdm
\Gamma(M^2) =
 2 P \int^{W^2}_{0} ds_2 \int^{W^2}_{0} ds_1
\frac{\rho(s_1,s_2)}{s_1 -s_2} e^{-s_2/M^2}
\edm
\be
= 2 b \int^{W^2}_{0} sds
e^{-s/M^2}\left[ W^2 + s\ln \frac{W^2}{s} \right]
\ee
Eq. (27) essentially differs from the corresponding expression for bare loop
contribution in ref. [12]: e.g. the integrand in (27) is positive. while in
[12] it is negative in the main region of integration. Of course, the QCD sum
rule calculation in this case has a serious drawback: the dependence of the
result on the continuum threshold $W^2$ is not in the form of a small
correction of the type $\exp (-W^2/M^2) \ll 1$ at $W^2 \gg M^2$, but much more
strong. This is a direct consequence of high (equal to 5) dimension of the
operator $U_\mu$. It is clear that the higher the dimension of the considered
operator is, the stronger will be the dependence on the continuum threshold and
less certain the results of QCD sum rules calculations. It must be stressed
that for operators of high dimensions the loop diagrams are in principle
nonrenormalizable, the role of excited states in the physical spectrum
increases and the determination of the lowest state contribution becomes
impossible.

Eq. (27) must be taken instead of the contribution of the bare loop diagram,
used in [12]. A similar procedure must be also applied in the case of other
terms in the sum rules [12], containing ultraviolet cut off. The results of
such calculation show, however, that occasionally the numerical value of $\ll
U\gg$ obtained in this way does not differ essentially form the value found in
[12] -- it is in the limit of quoted errors. (The same refers to the other
calculated in [12] matrix element $\ll V \gg$, related to the integral from
the
structure function $g_2(x)$.). Unfortunately, this fact does not mean that the
value of $\ll U \gg$ determined in [12] is reliable. The value $\ll U \gg$
obtained from the sum rule almost entirely comes from the last accounted in
[12] term of OPE of dimension 8. (The contribution of the bare loop
corresponds
to the background term -- the second term in eq. (15).) For this reason unless
the next term of OPE will be calculated in this problem, there is no confidence
in the result.

\section{Conclusions}

It was demonstrated, that in the QCD sum rules determination of hadronic
coupling constants with external fields the phenomenological sides of the sum
rules were not treated properly in some cases. The most suitable form of the
representation of the physical states contributions -- eq. (15) is suggested.
At the same time in the loop diagrams in the QCD sides of the sum rules the
integration in the double dispersion representation must be performed in the
intervals $0 < s_, s_2 < W^2$, where $W^2$ is the continuum threshold. It is
shown that for high (negative in mass) dimension of external field the results
of coupling constants determination in essential way depend on the value of the
continuum threshold $W^2$ and are less certain than other QCD sum rule results.
It was stressed the necessity the account in the nondiagonal transition terms,
corresponding to diagram of Fig. 4, not only nonsuppressed by Borel
transformation terms, but also the terms, exponentially suppressed by Borel
transformation -- the last term in eq. (15). The role of these nonaccounted in
previous calculation terms was estimated for few examples.

{\bf Acknowledgments.}
I am indepted to H. Leutwyler for the hospitality at Bern University, where
this work was finished and to A.V. Smilga for useful discussions. I am
thankful to O. H\"{a}nni for preparing the manuscript. This work
was supported in part by Schweizerischer Nationalfonds and by the INTAS Grant
93-283.

	 \newpage

\noindent {\bf Figure captions}

\hspace{1cm}

\noindent {\bf Fig. 1.} The bare loop diagram, corresonding to the
determination of
nucleon magnetic moments or to the quark mass correction to the baryon masses.
The solid lines correspond to quark propagators, crosses mean the action of
currents $\eta, \bar{\eta}$, the bubble corresponds to quark interaction with
external field.

\hspace{4mm}

\noindent {\bf Fig. 2.} The bare loop diagram for twist 4 correction to the
Bjorken sum rule for deep inelastic electron-nucleon scattering. The dashed
vertical line corresponds to discontinuity over $p^2_1$ at $p^2_1 \neq p^2_2$.

\hspace{4mm}

\noindent {\bf Fig. 3.} The integration domains in $s_1, s_2$ plane.

\hspace{4mm}

\noindent {\bf Fig. 4.} The schematical representation of $h \rightarrow
h^\star (h^\star \rightarrow h)$ transitions in the external field.

\hspace{4mm}

\noindent {\bf Fig. 5.} The diagram representing the contribution of the
operator $m_q \bar{q}q$ to the sum rule for quark mass correction to vector
meson mass.

\end{document}